\documentclass[twocolumn,aps,prc,superscriptaddress,showpacs,floatfix]{revtex4}
\usepackage{amssymb}
\usepackage{amsmath}
\usepackage{graphicx}
\usepackage{longtable}
\usepackage[normalem]{ulem}
\usepackage[dvips]{color}

\renewcommand{\vector}[1]{\boldsymbol{#1}}

\begin{document}

\title{Lorentz covariant nucleon self-energy decomposition of the nuclear
symmetry energy}
\author{Bao-Jun Cai}
\affiliation{Department of Physics, Shanghai Jiao Tong University, Shanghai 200240, China}
\author{Lie-Wen Chen\footnote{%
Corresponding author (email: lwchen$@$sjtu.edu.cn)}}
\affiliation{Department of Physics, Shanghai Jiao Tong University, Shanghai 200240, China}
\affiliation{Center of Theoretical Nuclear Physics, National Laboratory of Heavy-Ion
Accelerator, Lanzhou, 730000, China}
\date{\today}

\begin{abstract}
Using the Hugenholtz-Van Hove theorem, we derive analytical
expressions for the nuclear symmetry energy $E_{\text{sym}}(\rho )$
and its density slope $L(\rho )$ in terms of the Lorentz covariant
nucleon self-energies in isospin asymmetric nuclear matter. These
general expressions are useful for determining the density
dependence of the symmetry energy and understanding the Lorentz
structure and the microscopic origin of the symmetry energy in
relativistic covariant formulism. As an example, we analyze the
Lorentz covariant nucleon self-energy decomposition of
$E_{\text{sym}}(\rho )$ and $L(\rho )$ and derive the corresponding
analytical expressions within the nonlinear
$\sigma$-$\omega$-$\rho$-$\delta$ relativistic mean field model.
\end{abstract}

\maketitle

\section{Introduction}

In the current research of nuclear physics and astrophysics, there
is of great interest to study the density dependence of the nuclear
symmetry energy $E_{\text{sym}}(\rho )$ that essentially
characterizes the isospin dependent part of the equation of state
(EOS) of asymmetric nuclear matter. The exact knowledge on the
symmetry energy is important for understanding not only many
problems in nuclear physics, but also many critical topics in
astrophysics~\cite{LiBA98,Dan02,Lat04,Ste05,Bar05,LCK08} as well as
some interesting issues regarding possible new physics beyond the
standard model \cite{Hor01b,Sil05,Wen09}. During the last decade,
although significant progress has been made both experimentally and
theoretically on constraining the density dependence of the symmetry
energy~\cite{Bar05,LCK08,Che05,Tsa09,Cen09,Nat10} (see, e.g., Refs.
\cite{XuC10,Che10,Tsa11,Che11a,New11} for review of recent
progress), large uncertainties on $E_{\text{sym}}(\rho )$ still
exist, especially its super-normal density behavior remains elusive
and largely controversial~\cite{Xia09,Fen10,Rus11,XuC10b}. To reduce
the uncertainties of the constraints on $E_{\text{sym}}(\rho )$ is
thus of critical importance and remains a big challenge in the
community, and this provides a strong motivation for studying
isospin nuclear physics in radioactive nuclei at the new/planning
rare isotope beam facilities around the world, such as CSR/Lanzhou
and BRIF-II/Beijing in China, RIBF/RIKEN in Japan, SPIRAL2/GANIL in
France, FAIR/GSI in Germany, FRIB/NSCL in USA, SPES/LNL in Italy,
and KoRIA in Korea.

Theoretically, the uncertainties of the constraints on $E_{\text{sym}}(\rho
) $ are mainly due to the lack of knowledge about the isospin dependence of
in-medium nuclear effective interactions and the limitations in the
techniques for solving the nuclear many body problem. Very recently, it has
been proposed that it is very useful to directly decompose $E_{\text{sym}%
}(\rho )$ in terms of some relevant parts of the commonly used underlying
nuclear effective interaction in nuclear medium~\cite{XuC10,XuC11,Che12}.
Based on the Hugenholtz-Van Hove (HVH) theorem, indeed, the $E_{\text{sym}%
}(\rho )$ can be decomposed analytically in terms of the
single-nucleon potential in asymmetric nuclear matter and the
resulting expressions are quite general and independent of the
detailed nature of the nucleon interactions, providing an important
and physically more transparent approach to extract information on
the symmetry energy from the isospin dependence of strong
interaction in nuclear medium and understand why the symmetry energy
predicted from various models is so uncertain~\cite{XuC10b,Che12}.
In these works, the decomposition of $E_{\text{sym}}(\rho )$ is
based on non-relativistic framework. It is thus of great interest to
explore more general decomposition within relativistic covariant
framework, which is the main motivation of the present work.

\section{Covariant self-energy decomposition of $E_{\mathrm{%
sym}}(\protect\rho )$ and $L(\protect\rho )$}

The relativistic covariant formulation has made great success during
the last decades in understanding many nuclear
phenomena~\cite{Ser86,Ser97,Men06}. In particular, the microscopic
relativistic covariant Dirac-Brueckner-Hartree-Fock (DBHF)
approach~\cite{Ana83,Hor87,Bro90,Dal07,Dal10,Sam10} has achieved
impressive success in describing the saturation properties of
nuclear matter without any need to introduce a three-nucleon force
required in the microscopic non-relativistic BHF calculations (see,
e.g., Refs.~\cite{LiZH06,LiZH08}). It has been argued that in
non-relativistic calculations the three-nucleon forces must be
introduced to mimic the variation of the Dirac spinors in the
nuclear medium contained in relativistic covariant
approach~\cite{Dal10}. In addition, the Lorentz covariant
decomposition of the nuclear mean field potential has been shown to
be very important for understanding the dynamics in heavy ion
collisions at relativistic energies~\cite{Ko87,Bla93,Gai04}. These
features imply that the Lorentz covariance could be important for
understanding the higher energy/density nuclear phenomena, e.g., the
high density behaviors of the symmetry energy.

Owing to the translational, rotational and time-reversal invariance,
parity conservation, and hermiticity, the Lorentz covariant nucleon
self-energy in the rest frame of asymmetric nuclear matter with
baryon density $\rho =\rho _{\text{n}}+\rho _{\text{p}}$ and isospin
asymmetry $\alpha =(\rho _{\text{n}}-\rho _{\text{p}})/\rho $ can be
written generally as~\cite{Jam81,Hor83,Hor87,Ser86,Uec90},
\begin{align}
\Sigma ^{J}(\rho ,\alpha ,|\mathbf{k}|)=& \Sigma
_{\text{S}}^{J}(\rho
,\alpha ,|\mathbf{k}|)-\gamma _{\mu }\Sigma ^{\mu ,J}(\rho ,\alpha ,|\mathbf{%
k}|)  \notag  \\
=& \Sigma _{\text{S}}^{J}(\rho ,\alpha ,|\mathbf{k}|)+\gamma ^{0}\Sigma _{%
\text{V}}^{J}(\rho ,\alpha ,|\mathbf{k}|)  \notag \\
+&\vector{\gamma}\cdot \mathbf{k}^{0}\Sigma _{\text{K}}^{J}(\rho ,\alpha ,|%
\mathbf{k}|),  \label{DefSE2}
\end{align}%
where $J=\textrm{n}~\text{or}~\textrm{p}$ is isospin index; $\Sigma
_{\text{S}}^{J}(\rho ,\alpha ,|\mathbf{k}|)$ is the scalar
self-energy, $\Sigma _{\text{V}}^{J}(\rho ,\alpha
,|\mathbf{k}|)\equiv -\Sigma ^{0,J}(\rho ,\alpha ,|\mathbf{k}|)$ and
$\Sigma _{\text{K}}^{J}(\rho,\alpha ,|\mathbf{k}|)$ are,
respectively, the zeroth component and the space component of the
vector self-energy $\Sigma ^{\mu ,J}(\rho ,\alpha ,|\mathbf{k}|)$,
and they all generally depend on $\rho $, $\alpha $ and the
magnitude of the nucleon momentum $|\mathbf{k}|$ (Minkowski metric
is $g_{\mu \nu }=(+,-,-,-)$);
$\mathbf{k}^{0}=\mathbf{k}/|\mathbf{k}|$ is the unit vector of
momentum $\mathbf{k}$. A proof of Eq. (\ref{DefSE2}) can be found in
Ref.~\cite{Ser86}. Accordingly, the single-nucleon energy can be
expressed as~\cite{Jam81,Hor83,Hor87,Ser86,Uec90}
\begin{equation}
\mathcal{E}^{J}(\rho ,\alpha ,|\mathbf{k}|)=\mathcal{E}^{\ast
J}(\rho ,\alpha ,|\mathbf{k}|)+\Sigma _{\text{V}}^{J}(\rho ,\alpha
,|\mathbf{k}|), \label{DisRelaANM}
\end{equation}%
with
\begin{equation}
\mathcal{E}^{\ast J}(\rho ,\alpha ,|\mathbf{k}|)=\sqrt{\left[ \mathbf{k}%
^{\ast J}(\rho ,\alpha ,|\mathbf{k}|)\right] ^{2}+\left[ M_{J}^{\ast
}(\rho ,\alpha ,|\mathbf{k}|)\right] ^{2}},
\end{equation}%
where the nucleon effective (Dirac) mass $M_{J}^{\ast }(\rho ,\alpha
,|\mathbf{k}|)$ and effective momentum $\mathbf{k}^{J\ast }(\rho
,\alpha ,|\mathbf{k}|)$
are defined, respectively, as%
\begin{eqnarray}
M_{J}^{\ast }(\rho ,\alpha ,|\mathbf{k}|) &=&M+\Sigma
_{\text{S}}^{J}(\rho
,\alpha ,|\mathbf{k}|),~~ \\
\mathbf{k}^{\ast J}(\rho ,\alpha ,|\mathbf{k}|) &=&\mathbf{k}+\mathbf{k}%
^{0}\Sigma _{\text{K}}^{J}(\rho ,\alpha ,|\mathbf{k}|),~~
\end{eqnarray}%
with $M$ being the nucleon mass.

Due to the exchange symmetry between protons and neutrons in nuclear
matter, the EOS of asymmetric nuclear matter, defined by its binding
energy per nucleon, can be expanded as a power series of even-order
terms in $\alpha $ as
\begin{equation}
E(\rho ,\alpha )\simeq E_{0}(\rho )+E_{\text{sym}}(\rho )\alpha ^{2}+%
\mathcal{O}(\alpha ^{4}),  \label{EoSpert}
\end{equation}%
where $E_{0}(\rho )=E(\rho ,\alpha =0)$ is the EOS of symmetric
nuclear matter, and the symmetry energy is expressed as
\begin{equation}
E_{\text{sym}}(\rho )= \left. \frac{1}{2}\frac{\partial ^{2}E(\rho
,\alpha )}{\partial \alpha ^{2}}\right\vert _{\alpha =0}.
\label{DefEsym}
\end{equation}%
Around the nuclear matter saturation density $\rho _{0}$, the
symmetry energy can be expanded as
\begin{equation}
E_{\text{sym}}(\rho )\simeq E_{\text{sym}}(\rho _{0})+L\chi +\mathcal{O}%
(\chi ^{2}),  \label{Esympert}
\end{equation}%
where $\chi =(\rho -\rho _{0})/3\rho _{0}$ is a dimensionless
variable, and $L=L(\rho _{0})$ is the density slope parameter of the
symmetry energy at $\rho _{0}$ and thus carries important
information on the symmetry energy at both high and low densities.
Generally, the slope parameter of the symmetry energy at arbitrary
density $\rho $ is defined as
\begin{equation}
L(\rho )= 3\rho \frac{dE_{\text{sym}}(\rho )}{d\rho }. \label{DefL}
\end{equation}%

According to the HVH theorem \cite{Hug58,Sat99}, the nucleon
chemical potential in asymmetric nuclear matter should be equal to
its Fermi energy (the single-nucleon energy at Fermi surface), i.e.,
\begin{equation}  \label{HVHANM}
\mathcal{E}_{\text{F}}^{J}(\rho ,\alpha,k_{\text{F}}^{J})=\frac{\partial[%
\rho E(\rho,\alpha)]}{\partial \rho _{J}}+M,
\end{equation}
where $\mathcal{E}_{\text{F}}^{J}(\rho ,\alpha,k_{\text{F}}^{J})\equiv%
\mathcal{E}^{J}(\rho ,\alpha,k_{\text{F}}^{J})$ is the nucleon Fermi
energy, and $k_{\text{F}}^J=k_{\text{F}}(1+\tau_3^J\alpha)^{1/3}$
(we assume $\tau _{3}^{\text{n}}=1$ and $\tau _{3}^{\text{p}}=-1$ in
this work) is the nucleon Fermi momentum with $k_\text{F}=(3\pi
^{2}\rho /2)^{1/3}$ being the Fermi momentum in symmetric nuclear
matter at density $\rho $. It should be noted that the HVH theorem
is independent of the detailed nature of the interactions used and
is valid for any interacting self-bound infinite Fermi system, such
as infinite nuclear matter~\cite{Hug58,Sat99,Uec90}.

Expanding $E(\rho,\alpha)$ as a power series of even-order terms in
$\alpha $ on the right-hand side of Eq. (\ref{HVHANM}), we can
obtain
\begin{align}
\sum_{J=\text{p,n}}\tau_3^J\mathcal{E}_{\text{F}}^J(\rho,\alpha,k_{\text{F}%
}^J) =&\sum_{i=1}4iE_{\text{sym},2i}(\rho)\alpha^{2i-1},  \label{DetEsym} \\
\sum_{J=\text{p,n}}\mathcal{E}_{\text{F}}^J(\rho,\alpha,k_{\text{F}}^J)=&2%
\frac{\partial[\rho E_0(\rho)]}{\partial\rho}+2M  \notag \\
+&2\frac{\partial}{\partial\rho}\sum_{i=1}\rho E_{\text{sym}%
,2i}(\rho)\alpha^{2i},  \label{DetL}
\end{align}
where $E_{\text{sym},2i}(\rho)\equiv [{1}/{(2i)!}][{\partial
^{2i}E(\rho ,\alpha )}/{\partial \alpha ^{2i}}]_{\alpha =0}$ are the
symmetry energies of different orders and particularly we have
$E_{\text{sym},2}(\rho)\equiv E_{\text{sym}}(\rho)$. Furthermore,
expanding $\mathcal{E}_{\text{F}}^{J}(\rho
,\alpha,k_{\text{F}}^{J})$ as a power series of $\alpha$ on the
left-hand side of Eq. (\ref{DetEsym}) and Eq. (\ref{DetL}), and
comparing the coefficients of the first-order $\alpha$ terms on both
left- and right-hand sides of Eq. (\ref{DetEsym}), we then obtain
\begin{align}
E_{\text{sym}}(\rho)=&\left.\frac{1}{4}\frac{d}{d\alpha} \left[%
\sum_{J=\text{p,n}}\tau_3^J\mathcal{E}_{\text{F}}^J(\rho,\alpha,k_{\text{F}%
}^J) \right]\right|_{\alpha=0},  \label{ForEsym}
\end{align}
while comparing the coefficients of second-order $\alpha$ terms on
both sides of Eq. (\ref{DetL}) leads to the following expression:
\begin{align}
L(\rho)=&\left.\frac{3}{4}\frac{d^2}{d\alpha^2} \left[\sum_{J=%
\text{p,n}}\mathcal{E}_{\text{F}}^J(\rho,\alpha,k_{\text{F}}^J) \right]%
\right|_{\alpha=0} +3E_{\text{sym}}(\rho).  \label{ForL}
\end{align}
Substituting Eq. (\ref{DisRelaANM}) into Eq. (\ref{ForEsym}), we can
obtain
\begin{align}
E_{\mathrm{sym}}(\rho )=&E_{\mathrm{sym}}^{\mathrm{kin}}(\rho )  \notag \\
+&E_{\mathrm{sym}}^{\mathrm{0,mom,K}}(\rho)+ E_{\mathrm{sym}}^{\mathrm{0,mom,S}%
}(\rho)+E_{\mathrm{sym}}^{\mathrm{0,mom,V}}(\rho)  \notag \\
+&E_{\mathrm{sym}}^{\mathrm{1st,K}}(\rho) +E_{\mathrm{sym}}^{\mathrm{1st,S}%
}(\rho) +E_{\mathrm{sym}}^{\mathrm{1st,V}}(\rho),  \label{ExprEsym}
\end{align}
where $E_{\mathrm{sym}}^{\mathrm{kin}}(\rho)$, $E_{\mathrm{sym}}^{\mathrm{0,mom},%
\mathcal{O} }(\rho )$ and
$E_{\mathrm{sym}}^{\mathrm{1st},\mathcal{O}}(\rho ) $ (here
$\mathcal{O}$ denotes K, S or V) represent, respectively, the
contributions from the kinetic part, the momentum dependence of the
nucleon self-energies in symmetric nuclear matter and the
first-order symmetry self-energies, and they can be expressed
analytically as
\begin{align}
E_{\text{sym}}^{\text{kin}}(\rho)=&\frac{k_{\text{F}}k_{\text{F}}^{\ast}(\rho)}{6
\mathcal{E}_{\text{F}}^{\ast}(\rho)},  \label{DefEsymkin} \\
E_{\mathrm{sym}}^{\mathrm{0,mom,K}}(\rho)=&
\frac{k_{\text{F}}k_{\text{F}}^{\ast}(\rho)}{6\mathcal{E}_{\text{F}}^{\ast}(\rho)}
\left.\frac{\partial \Sigma_{\text{K}}^0(\rho
,|\mathbf{k}|)}{\partial|\mathbf{k}|}\right|_{|\mathbf{k}|=k_{\text{F}}},
\label{DefEsymmomK} \\
E_{\mathrm{sym}}^{\mathrm{0,mom,S}}(\rho)=&
\frac{k_{\text{F}}M_0^{\ast}(\rho)}{6\mathcal{E}_{\text{F}}^{\ast}(\rho)}
\left.\frac{\partial \Sigma_{\text{S}}^0(\rho
,|\mathbf{k}|)}{\partial|\mathbf{k}|}\right|_{|\mathbf{k}|=k_{\text{F}}},
\label{DefEsymmomS} \\
E_{\mathrm{sym}}^{\mathrm{0,mom,V}}(\rho)=& \frac{k_{\text{F}}}{6}
\left.\frac{\partial \Sigma_{\text{V}}^0(\rho ,|\mathbf{k}|)}
{\partial|\mathbf{k}|}\right|_{|\mathbf{k}|=k_{\text{F}}},  \label{DefEsymmomV} \\
E_{\mathrm{sym}}^{\mathrm{1st,K}}(\rho)=&\frac{1}{2}\frac{k_{\text{F}%
}^{\ast}(\rho) \Sigma_{\text{K}}^{\mathrm{sym,1}}(\rho
,|\mathbf{k}|=k_{\text{F}})}{\mathcal{E}_{\text{F}}^{\ast}(\rho)},
\label{DefEsym1stK} \\
E_{\mathrm{sym}}^{\mathrm{1st,S}}(\rho)=&\frac{1}{2}\frac{M_0^{\ast}(\rho) \Sigma_{%
\text{S}}^{\mathrm{sym,1}}(\rho
,|\mathbf{k}|=k_{\text{F}})}{\mathcal{E}_{\text{F}}^{\ast}(\rho)},
\label{DefEsym1stS} \\
E_{\mathrm{sym}}^{\mathrm{1st,V}}(\rho)=&\frac{1}{2}\Sigma_{\text{V}}^{%
\mathrm{sym,1}}(\rho ,|\mathbf{k}|=k_{\text{F}}) ,
\label{DefEsym1stV}
\end{align}%
where $k_{\text{F}}^{\ast }(\rho)=k_{\text{F}}+\Sigma
_{\text{K}}^{0}(\rho ,k_{\text{F}})$, $M_{0}^{\ast }(\rho)=M+\Sigma
_{\text{S}}^{0}(\rho ,k_{\text{F}})$, $\mathcal{E}_{\text{F}}^{\ast
}(\rho)=({k_{\text{F}}^{\ast 2}+M_{0}^{\ast 2}})^{1/2}$, $\Sigma
_{\text{K}}^{0}(\rho ,|\mathbf{k}|)=\Sigma _{\text{K}}^{J}(\rho
,\alpha =0,|\mathbf{k}|)$, $\Sigma _{\text{S}}^{0}(\rho
,|\mathbf{k}|)=\Sigma _{\text{S}}^{J}(\rho ,\alpha
=0,|\mathbf{k}|)$, $\Sigma _{\text{V}}^{0}(\rho
,|\mathbf{k}|)=\Sigma _{\text{V}}^{J}(\rho ,\alpha
=0,|\mathbf{k}|)$, and the $i$-th order symmetry self-energy is
defined as (here $\mathcal{O}=\text{K,S,V}$)
\begin{equation}
\Sigma _{\mathcal{O}}^{\mathrm{sym},i}(\rho ,|\mathbf{k}|)=\left. \frac{1}{i!%
}\frac{\partial ^{i}}{\partial \alpha ^{i}}\left[ \sum_{J=\text{p,n}}\frac{%
\tau _{3}^{Ji}\Sigma _{\mathcal{O}}^{J}(\rho ,\alpha ,|\mathbf{k}|)}{2}%
\right] \right\vert _{\alpha =0}.  \label{SymmSE}
\end{equation}
Furthermore, Eq. (\ref{ExprEsym}) can be rewritten as
\begin{align}
E_{\mathrm{sym}}(\rho)&=\left.\frac{|\mathbf{k}|^2}{6M^{\ast}_{0,\textrm{Lan}}(\rho
,|\mathbf{k}|)}
\right|_{|\mathbf{k}|=k_{\text{F}}}  \notag \\
&+E_{\mathrm{sym}}^{\mathrm{1st,K}}(\rho) +E_{\mathrm{sym}}^{\mathrm{1st,S}%
}(\rho) +E_{\mathrm{sym}}^{\mathrm{1st,V}}(\rho),
\label{EsymDecomp}
\end{align}
where $M^{\ast}_{0,\textrm{Lan}}(\rho ,|\mathbf{k}|)$ is the nucleon
Landau mass in symmetric nuclear matter, i.e.,
$M^{\ast}_{0,\textrm{Lan}}(\rho
,|\mathbf{k}|)=|\mathbf{k}|[{d|\mathbf{k}|}/{d\mathcal{E}^{0}(\rho
,|\mathbf{k}|)}]^{-1}$ (see, e.g., Ref.~\cite{Che07}) with
$\mathcal{E}^{0}(\rho ,|\mathbf{k}|)= \mathcal{E}^{J}(\rho ,\alpha
=0,|\mathbf{k}|)$, and one can easily verify the relation
${k_{\text{F}}^2}/{6M^{\ast}_{0,\textrm{Lan}}(\rho
,k_{\text{F}})}=E_{\textrm{sym}}^{\textrm{kin}}(\rho)+
E_{\mathrm{sym}}^{\mathrm{0,mom,K}}(\rho)+
E_{\mathrm{sym}}^{\mathrm{0,mom,S}}(\rho)+
E_{\mathrm{sym}}^{\mathrm{0,mom,V}}(\rho)$. In this way, we have
decomposed analytically the symmetry energy $E_{\mathrm{sym}}(\rho
)$ in terms of the Lorentz covariant nucleon self-energies in
asymmetric nuclear matter.

Similarly, by substituting Eq. (\ref{DisRelaANM}) into Eq.
(\ref{ForL}), the slope parameter $L(\rho)$ can be decomposed as
\begin{align}  \label{ExprL}
L(\rho)=&L^{\mathrm{kin}}(\rho)+L^{\mathrm{mom}}(\rho)
+L^{\mathrm{1st}}(\rho)  \notag \\
&+ L^{\mathrm{cross}}(\rho) +L^{\mathrm{2nd}}(\rho),
\end{align}
with
\begin{align}
&L^{\mathrm{kin}}(\rho)=\frac{k_{\text{F}}k_{\text{F}}^{\ast}}{6\mathcal{E}_{\text{F}}^{\ast}}
+\frac{k_{\text{F}}^2M_0^{\ast2}}{6\mathcal{E}_{\text{F}}^{\ast3}},
\label{DefLKin}
\end{align}
\begin{align}
&L^{\mathrm{mom}}(\rho)=\frac{k_{\text{F}}^2M_0^{\ast2}}
{3\mathcal{E}_{\text{F}}^{\ast2}}\left.\frac{\partial\Sigma_{\text{K}}^0}{\partial
|\mathbf{k}|}\right|_{|\mathbf{k}|=k_{\text{F}}}  \notag \\
&+\frac{k_{\text{F}}^2}{6}\left[\frac{k_{\text{F}}^{\ast}}{\mathcal{E}_{%
\text{F}}^{\ast}} \frac{\partial^2\Sigma_{\text{K}}^0}{\partial |\mathbf{k}%
|^2}+\frac{M_0^{\ast}}{\mathcal{E}_{\text{F}}^{\ast}} \frac{%
\partial^2\Sigma_{\text{S}}^0}{\partial |\mathbf{k}|^2}+\frac{%
\partial^2\Sigma_{\text{V}}^0}{\partial |\mathbf{k}|^2}\right]_{|\mathbf{k}%
|=k_{\text{F}}}  \notag \\
&+\frac{k_{\text{F}}}{6}\left[\frac{k_{\text{F}}^{\ast}}{\mathcal{E}_{\text{F%
}}^{\ast}} \frac{\partial\Sigma_{\text{K}}^0}{ \partial|\mathbf{k}|} +\frac{%
M_0^{\ast}}{\mathcal{E}_{\text{F}}^{\ast}}\frac{\partial\Sigma_{\text{S}}^0}{%
\partial|\mathbf{k}|} +\frac{\partial\Sigma_{\text{V}}^0}{ \partial|\mathbf{k%
}|}\right]_{|\mathbf{k}|=k_{\text{F}}}  \notag \\
&+\frac{k_{\text{F}}^2}{6\mathcal{E}_{\text{F}}^{\ast3}}\left[%
M_0^{\ast2}\left( \frac{\partial\Sigma_{\text{K}}^0}{\partial |\mathbf{k}|}%
\right)^2+k_{\text{F}}^{\ast2}\left(\frac{\partial\Sigma_{\text{S}}^0}
{\partial |\mathbf{k}|}\right)^2\right]_{|\mathbf{k}|=k_{\text{F}}}  \notag \\
&-\frac{k_{\text{F}}^2k_{\text{F}}^{\ast}M_0^{\ast}}
{3\mathcal{E}_{\text{F}}^{\ast3}}\left[
\frac{\partial\Sigma_{\text{S}}^0}{\partial
|\mathbf{k}|}\left(1+\frac{\partial\Sigma_{\text{K}}^0}{\partial
|\mathbf{k}|}\right)\right]_{|\mathbf{k}|=k_{\text{F}}} ,
\label{DefLMDen} \\
&L^{\mathrm{1st}}(\rho)=\frac{3}{2\mathcal{E}_{\text{F}}^{\ast3}}\left[%
M_0^{\ast}\Sigma_{\text{K}}^{\mathrm{{sym},1}}- k_{\text{F}}^{\ast}\Sigma_{%
\text{S}}^{\mathrm{{sym},1}}\right]^2  \notag \\
&+\frac{3}{2}\left[\frac{k_{\text{F}}^{\ast}}{\mathcal{E}_{\text{F}}^{\ast}}%
\Sigma_{\text{K}}^{\mathrm{{sym},1}}+ \frac{M_0^{\ast}}{\mathcal{E}_{\text{F}%
}^{\ast}}\Sigma_{\text{S}}^{\mathrm{{sym},1}} +\Sigma_{\text{V}}^{\mathrm{{%
sym},1}}\right]  \notag \\
&+\frac{k_{\text{F}}M_0^{\ast2}\Sigma_{\text{K}}^{\text{sym,1}}}{ \mathcal{E}%
_{\text{F}}^{\ast3}} -\frac{k_{\text{F}}k_{\text{F}}^{\ast}M_0^{\ast}\Sigma_{%
\text{S}}^{\text{sym,1}}}{ \mathcal{E}_{\text{F}}^{\ast3}},
\label{DefLS1} \\
&L^{\text{cross}}(\rho)=  \notag \\
&k_{\text{F}}\left[\frac{k_{\text{F}}^{\ast}}{\mathcal{E}_{\text{F}}^{\ast}}%
\frac{\partial \Sigma_{\text{K}}^{\mathrm{{sym},1}}}{\partial |\mathbf{k}|}+%
\frac{M_0^{\ast}}{\mathcal{E}_{\text{F}}^{\ast}}\frac{\partial \Sigma_{\text{%
S}}^{\mathrm{{sym},1}}}{\partial|\mathbf{k}|}+\frac{\partial \Sigma_{\text{V}%
}^{\mathrm{{sym},1}}}{\partial |\mathbf{k}|}\right]_{|\mathbf{k}|=k_{\text{F}%
}}  \notag \\
&-\frac{k_{\text{F}}\Sigma_{\text{K}}^{\mathrm{{sym},1}}}{\mathcal{E}_{\text{%
F}}^{\ast}} \left[\frac{k_{\text{F}}^{\ast2}}{\mathcal{E}_{\text{F}}^{\ast2}}%
\left(\frac{\partial\Sigma_{\text{K}}^0}{ \partial|\mathbf{k}|}+\frac{%
M_0^{\ast}}{k_{\text{F}}^{\ast}}\frac{\partial\Sigma_{\text{S}}^0}{ \partial|%
\mathbf{k}|}\right) -\frac{\partial\Sigma_{\text{K}}^0}{ \partial|\mathbf{k}|%
}\right]_{|\mathbf{k}|=k_{\text{F}}}  \notag \\
&-\frac{k_{\text{F}}\Sigma_{\text{S}}^{\mathrm{{sym},1}}}{\mathcal{E}_{\text{%
F}}^{\ast}} \left[ \frac{M_0^{\ast2}}{\mathcal{E}_{\text{F}}^{\ast2}}\left(%
\frac{k_{\text{F}}^{\ast}}{M_0^{\ast}} \frac{\partial\Sigma_{\text{K}}^0}{
\partial|\mathbf{k}|}+\frac{\partial\Sigma_{\text{S}}^0}{ \partial|\mathbf{k}%
|}\right)-\frac{ \partial\Sigma_{\text{S}}^0}{ \partial|\mathbf{k}|}\right]%
_{|\mathbf{k}|=k_{\text{F}}} ,  \label{DefLCross} \\
&L^{\mathrm{2nd}}(\rho)=3\left[\frac{k_{\text{F}}^{\ast}}{\mathcal{E}_{\text{%
F}}^{\ast}}\Sigma_{\text{K}}^{\mathrm{{sym},2}} +\frac{M_0^{\ast}}{\mathcal{E%
}_{\text{F}}^{\ast}}\Sigma_{\text{S}}^{\mathrm{{sym},2}}+\Sigma_{\text{V}}^{%
\mathrm{{sym},2}}\right].  \label{DefLS2}
\end{align}
On the right-hand side of Eqs.~(\ref{DefLKin})-(\ref{DefLS2}), the
density and momentum dependence have been suppressed with $\Sigma
_{\mathcal{O}}^{\mathrm{sym},i}=\Sigma
_{\mathcal{O}}^{\mathrm{sym},i}(\rho ,|\mathbf{k}|=k_{\text{F}})$
($\mathcal{O}=\text{K,S,V}$). Eq. (\ref{ExprEsym}) (or
(\ref{EsymDecomp})) and Eq. (\ref{ExprL}) are two main results of
this work.

\section{Application to the nonlinear $\sigma$-$\omega$-$\rho$-$\delta$ RMF model}

The nucleon self-energies can be calculated theoretically from a
certain relativistic covariant approach or extracted experimentally
(around $\rho_0$) from the Dirac phenomenology of nucleon-nucleus
scattering. The Lorentz covariant nucleon self-energy decompositions
of $E_{\text{sym}}(\rho)$ in Eq. (\ref{ExprEsym}) (or
(\ref{EsymDecomp})) and $L(\rho)$ in Eq. (\ref{ExprL}) are general
and they are useful for determining the density dependence of the
symmetry energy and understanding its Lorentz structure and the
microscopic origin. As an example, we consider here the nonlinear
$\sigma$-$\omega$-$\rho$-$\delta$ relativistic mean field (RMF)
model which is based on effective interaction Lagrangians involving
nucleon and meson fields, and has been widely discussed in the
literature (see, e.g., Ref.~\cite{Che07}). A very useful feature of
this model is that the nucleon self-energies in asymmetric nuclear
matter can be obtained analytically and this makes our analysis
physically transparent. The Lagrangian density of the nonlinear
$\sigma$-$\omega$-$\rho$-$\delta$ RMF model can be expressed as
(see, e.g., Ref.~\cite{Che07}):
\begin{align}  \label{NonRMF}
\mathcal{L}=&\bar{\psi}\left[\gamma_{\mu}(i\partial^{\mu}-g_{\omega}\omega^{%
\mu})-(M-g_{\sigma}\sigma)\right]\psi  \notag \\
+&\frac{1}{2}\left(\partial_{\mu}\sigma\partial^{\mu}\sigma-m_{\sigma}^2%
\sigma^2\right) -\frac{1}{4}F_{\mu\nu}F^{\mu\nu}+\frac{1}{2}%
m_{\omega}^2\omega_{\mu}\omega^{\mu}  \notag \\
-&\frac{1}{3}b_{\sigma}M(g_{\sigma}\sigma)^3 -\frac{1}{4}c_{\sigma}(g_{%
\sigma}\sigma)^4+\frac{1}{4}c_{\omega}(g_{\omega}^2\omega_{\mu}\omega^{%
\mu})^2  \notag \\
+&\frac{1}{2}\left(\partial_{\mu}\vec{\mkern1mu\delta}\cdot\partial^{\mu}%
\vec{\mkern1mu\delta} -m_{\delta}^2\vec{\mkern1mu\delta}^2\right) +\frac{1}{2%
}m_{\rho}^2\vec{\mkern1mu\rho}_{\mu}\cdot\vec{\mkern1mu\rho}^{\mu} -\frac{1}{%
4}\vec{\mkern1mu G}_{\mu\nu}\cdot\vec{\mkern1mu G}^{\mu\nu}  \notag \\
+&\frac{1}{2}\left(g_{\rho}^2\vec{\mkern1mu\rho}_{\mu}\cdot\vec{\mkern1mu\rho%
}^{\mu}\right)\left(\Lambda_{\text{S}}g_{\sigma}^2\sigma^2+ \Lambda_{\text{V}%
}g_{\omega}^2\omega_{\mu}\omega^{\mu}\right)  \notag \\
-&g_{\rho}\vec{\mkern1mu\rho}_{\mu}\cdot\bar{\psi}\gamma^{\mu}\vec{\tau}%
^J\psi +g_{\delta}\vec{\mkern1mu\delta}\cdot\bar{\psi}\vec{\tau}^J\psi,
\end{align}
where $F_{\mu\nu}\equiv\partial_{\mu}\omega_{\nu}-\partial_{\nu}\omega_{\mu}$
and $\vec{\mkern1mu G}_{\mu\nu}\equiv\partial_{\mu}\vec{\mkern1mu\rho}%
_{\nu}-\partial_{\nu}\vec{\mkern1mu\rho}_{\mu} $ are strength
tensors of $\omega$ field and $\rho$ field, respectively while
$\psi$, $\sigma$, $\omega_{\mu}$, $\vec{\mkern1mu\rho}_{\mu}$ and
$\vec{\mkern1mu\delta}$ are nucleon field, isoscalar-scalar field,
isoscalar-vector field, isovector-vector and isovector-scalar field,
respectively, and the arrows denote the isospin vector.
$\Lambda_{\text{S}}$ and $\Lambda_{\text{V}}$ are two cross-coupling
constants for varying the density dependence of
$E_{\mathrm{sym}}(\rho )$, and $m_{\sigma}$, $m_{\omega}$,
$m_{\rho}$, and $m_{\delta}$ are masses of mesons.

In the RMF model, meson fields are replaced by their expectation
values, i.e., $\bar{\sigma}\rightarrow \sigma $,
$\bar{\omega}_{0}\rightarrow \omega _{\mu } $,
$\bar{\rho}_{0}^{(3)}\rightarrow \vec{\mkern1mu\rho }_{\mu }$, where
the subscript \textquotedblleft $0$" denotes the zeroth component of
the four-vector while the superscript \textquotedblleft ($3$)"
denotes the third component of isospin, Furthermore, the space-like
self-energy $\Sigma_{\text{K}}^J(\rho,\delta,|\mathbf{k}|)$ vanishes
(due to the Hartree approximation in the RMF model) while the scalar
and time-like self-energies are momentum independent, i.e.,
\begin{align}
\Sigma_{\text{S}}^J(\rho,\alpha)&=-g_{\sigma}\bar{\sigma}+\tau_3^Jg_{\delta}\bar{%
\delta}^{(3)},  \label{MeanSelfEnergy1} \\
\Sigma_{\text{V}}^J(\rho,\alpha)&=g_{\omega}\bar{\omega}_0-\tau_3^Jg_{\rho}\bar{\rho%
}_0^{(3)}.  \label{MeanSelfEnergy2}
\end{align}
The symmetry energy then can be decomposed as
\begin{align}
E_{\mathrm{sym}}(\rho )=& E_{\mathrm{sym}}^{\mathrm{kin}}(\rho )+E_{\mathrm{%
sym}}^{\mathrm{1st,S}}(\rho )+E_{\mathrm{sym}}^{\mathrm{1st,V}}(\rho
) \notag \\
=& \frac{k_{\text{F}}^{2}}{6\mathcal{E}_{\text{F}}^{\ast }}+\frac{1}{2}\frac{%
M_{0}^{\ast }\Sigma _{\text{S}}^{\mathrm{{sym},1}}(\rho )}{\mathcal{E}_{%
\text{F}}^{\ast }} +\frac{1}{2}\Sigma
_{\text{V}}^{\mathrm{{sym},1}}(\rho ), \label{RMFEsymDecom1}
\end{align}%
where the ($1$st-order) symmetry self-energies are
\begin{align}
\Sigma _{\text{S}}^{\mathrm{{sym},1}}(\rho )&=-\frac{g_{\delta
}^{2}M_{0}^{\ast }\rho }{\mathcal{E}_{\text{F}}^{\ast }Q_{\delta }}, \\
\Sigma _{\text{V}}^{\mathrm{{sym},1}}(\rho )&=+\frac{g_{\rho }^{2}\rho }{%
Q_{\rho }},
\end{align}
with $Q_{\delta }=m_{\delta }^{2}+3g_{\delta }^{2}(\rho _{\text{S}%
}/M_{0}^{\ast }-\rho /\mathcal{E}_{\text{F}}^{\ast })$, $\rho
_{\text{S}}$ being the scalar density, and $Q_{\rho }=m_{\rho
}^{2}+\Lambda _{\text{S}}g_{\sigma }^{2}g_{\rho
}^{2}\bar{\sigma}^{2}+\Lambda _{\text{V}}g_{\omega }^{2}g_{\rho
}^{2}\bar{\omega}_{0}^{2}$. We note that the above analytical
expression for $E_{\mathrm{sym}}(\rho )$ is exactly the same as the
one obtained from the normal approach (see, e.g.,
Ref.~\cite{Che07}). Similarly, the slope parameter $L(\rho)$ can be
decomposed as
\begin{equation}
L(\rho)=L^{\mathrm{kin}}(\rho)+L^{\mathrm{1st}}(\rho)+L^{\mathrm{2nd}}(\rho),
\label{RMFLDecom}
\end{equation}
with
\begin{align}
L^{\mathrm{kin}}(\rho)=&\frac{k_{\text{F}}^2(\mathcal{E}_{\text{F}%
}^{\ast2}+M_0^{\ast2})}{6\mathcal{E}_{\text{F}}^{\ast3}},
\label{RMFLDecomkin1} \\
L^{\mathrm{1st}}(\rho)=&\frac{3}{2}\left[\frac{M_0^{\ast}\Sigma_{\text{S}}^{%
\mathrm{{sym},1}}(\rho)}{\mathcal{E}_{\text{F}}^{\ast}} +\Sigma_{\text{V}}^{%
\mathrm{{sym},1}}(\rho)\right]  \notag \\
+&\frac{3k_{\text{F}}^2}{2\mathcal{E}_{\text{F}}^{\ast3}}\left[\Sigma_{\text{%
S}}^{\mathrm{{sym},1}}(\rho) \right]^2-\frac{M_0^{\ast}k_{\text{F}}^2\Sigma_{%
\text{S}}^{\mathrm{{sym},1}}(\rho)}{\mathcal{E}_{\text{F}}^{\ast3}},
\label{RMFLDecom1stsymm1} \\
L^{\mathrm{2nd}}(\rho)=&3\left[\frac{M_0^{\ast}\Sigma_{\text{S}}^{\mathrm{{%
sym},2}}(\rho)}{\mathcal{E}_{\text{F}}^{\ast}}+\Sigma_{\text{V}}^{\mathrm{{%
sym},2}}(\rho)\right],  \label{RMFLDecom2ndsymm1}
\end{align}
where the $2$nd-order symmetry self-energies are
\begin{align}
\Sigma_{\text{S}}^{\mathrm{{sym},2}}(\rho)=&-\frac{g_{\sigma}}{2Q_{\sigma}}%
\Bigg(\frac{g_{\delta}^2M_0^{\ast2}\rho\Gamma}{ \mathcal{E}_{\text{F}%
}^{\ast2}Q_{\delta}}
-\frac{2g_{\delta}^2M_0^{\ast}k_{\text{F}}^2\rho^2}{\mathcal{E}_{\text{F}
}^{\ast4}Q_{\delta}}  \notag \\
&-\frac{g_{\sigma}M_0^{\ast}k_{\text{F}}^2\rho}{3%
\mathcal{E}_{\text{F}}^{\ast3}} +\frac{2g_{\sigma}^2g_{\rho}^4\Lambda_{\text{%
S}}\bar{\sigma}\rho^2}{Q_{\rho}^2} \Bigg),  \label{RMFSE2S2} \\
\Sigma_{\text{V}}^{\mathrm{{sym},2}}(\rho)=&-\frac{g_{\omega}^3g_{\rho}^4%
\Lambda_{\text{V}}\bar{\omega}_0\rho^2}{Q_{\omega}Q_{\rho}^2},
\label{RMFSE2V1}
\end{align}
with $Q_{\sigma}=m_{\sigma}^2+g_{\sigma}^2({3\rho_{%
\text{S}}}/{{M_0^{\ast}}}-{3\rho}/{\mathcal{E}_{\text{F}}^{\ast}})
+2b_{\sigma}Mg_{\sigma}^3\bar{\sigma}+3c_{\sigma}g_{\sigma}^4\bar{\sigma}%
^2,Q_{\omega}=m_{\omega}^2+3c_{\omega}g_{\omega}^4\bar{\omega}_0^2 $ and $%
\Gamma=3g_{\sigma}g_{\delta}^2({2\rho_{\text{S}}}/{M_0^{\ast2}} -{3\rho}/{%
M_0^{\ast}\mathcal{E}_{\text{F}}^{\ast}}+{M_0^{\ast}\rho}/{\mathcal{E}_{%
\text{F}}^{\ast3}})$. To the best of our knowledge, the above
formulas give, for the first time, the analytical expression of the
slope parameter $L(\rho)$ in the nonlinear RMF model. The above
analytical expressions of $E_{\mathrm{sym}}(\rho )$ and $L(\rho)$
can be easily generalized to the case of the density dependent RMF
model that has similar isospin structure as the nonlinear RMF
model~\cite{Che07}. It should be mentioned that these analytical
expressions for $E_{\mathrm{sym}}(\rho )$ and $L(\rho)$ are very
useful for determining the isovector parameters in the RMF model by
fitting the empirical properties of asymmetric nuclear matter (see,
e.g., Ref.~\cite{Gle00} for such a procedure in the case of the
isoscalar sector).

\begin{figure}[h!]
\includegraphics[width=8.0cm]{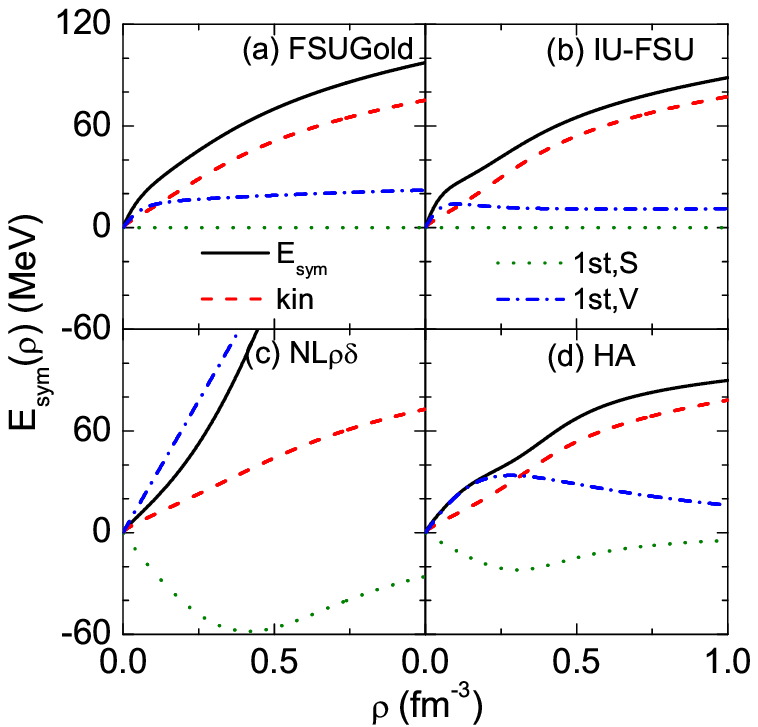}
\caption{(Color online) Density dependence of $E_{\mathrm{sym}}(\rho
)$ and its self-energy decomposition according to
Eq.~(\protect\ref{RMFEsymDecom1}) in the nonlinear RMF model with
different interactions.} \label{FIGEsymDecom}
\end{figure}

Shown in Fig.~\ref{FIGEsymDecom} is the density dependence of
$E_{\mathrm{sym}}(\rho )$ and its self-energy decomposition
according to Eq. (\ref{RMFEsymDecom1}) for four interactions, i.e.,
FSUGold~\cite{Tod05}, IU-FSU~\cite{Fat10},
NL$\rho\delta$~\cite{Liu02} and HA~\cite{Bun03}. FSUGold and IU-FSU
are two accurately calibrated interactions based on the ground state
properties of closed-shell nuclei, their linear response, and the
structure of neutron stars. Since FSUGold and IU-FSU do not consider
the isovector-scalar $\delta$ meson field, one thus has $\Sigma
_{\text{S}}^{\mathrm{{sym},1}}(\rho )=0$. On the other hand, both
NL$\rho\delta$ and HA include the $\delta$ meson field with the
latter fitting successfully some results obtained from the
microscopic DBHF approach while the former fitting the empirical
properties of asymmetric nuclear matter and describing reasonably
well the binding energies and charge radii of a large number of
nuclei~\cite{Gai04}. From Fig.~\ref{FIGEsymDecom}, one can see that
while the kinetic contribution
$E_{\mathrm{sym}}^{\mathrm{kin}}(\rho)$ is roughly the same for
different interactions, the different interactions predict
significantly different values for
$E_{\mathrm{sym}}^{\mathrm{1st,S}}(\rho )$ and
$E_{\mathrm{sym}}^{\mathrm{1st,V}}(\rho )$. In particular, compared
with FSUGold and IU-FSU, HA predicts very similar total
$E_{\mathrm{sym}}(\rho )$ but significantly different
$E_{\mathrm{sym}}^{\mathrm{1st,S}}(\rho )$ and
$E_{\mathrm{sym}}^{\mathrm{1st,V}}(\rho )$.

\begin{figure}[h!]
\includegraphics[width=8.0cm]{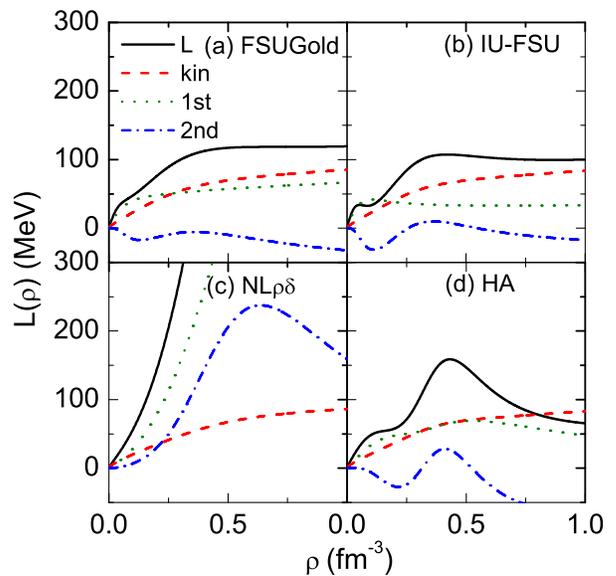}
\caption{(Color online) Density dependence of $L(\rho)$ and its
self-energy decomposition according to Eq.~(\protect\ref{RMFLDecom})
in the nonlinear RMF model with different interactions.}
\label{FIGLDecom}
\end{figure}

Similarly, we show in Fig.~\ref{FIGLDecom} the density dependence of
the slope parameter $L(\rho)$ and its self-energy decomposition
according to Eq. (\ref{RMFLDecom}) for FSUGold, IU-FSU,
NL$\rho\delta$ and HA. Again, it is seen that the different
interactions predict roughly same kinetic contribution
$L^{\mathrm{kin}}(\rho)$ but significantly different values for
$L^{\mathrm{1st}}(\rho)$ and $L^{\mathrm{2nd}}(\rho)$. In
particular, one can see that the higher-order contribution
$L^{\mathrm{2nd}}(\rho)$ from the second-order symmetry
self-energies generally cannot be neglected, agreeing well with the
recent non-relativistic calculations~\cite{Che12}.

\section{Summary and outlook}
Using the Hugenholtz-Van Hove theorem, we have shown that the
symmetry energy $E_{\text{sym}}(\rho )$ and its density slope
$L(\rho )$ can be decomposed analytically in terms of the Lorentz
covariant nucleon self-energies in asymmetric nuclear matter, and
the corresponding expressions have been derived for the first time.
These general expressions for the covariant self-energy
decomposition of $E_{\text{sym}}(\rho )$ and $L(\rho )$ are useful
for determining the density dependence of the symmetry energy,
deciphering the Lorentz structure of the symmetry energy, and
understanding the microscopic origins of the symmetry energy. As an
example, we have analyzed the Lorentz covariant nucleon self-energy
decomposition of $E_{\text{sym}}(\rho )$ and $L(\rho )$ within the
nonlinear $\sigma$-$\omega$-$\rho$-$\delta$ relativistic mean field
model and derived the corresponding analytical expressions for
$E_{\text{sym}}(\rho )$ and $L(\rho )$, which are potentially useful
for fixing the isovector parameters in the RMF model from fitting
the empirical properties of asymmetric nuclear matter.

From analyzing the self-energy decomposition of $E_{\text{sym}}(\rho
)$ and $L(\rho )$ within the nonlinear
$\sigma$-$\omega$-$\rho$-$\delta$ relativistic mean field model, we
have found that the results strongly depend on the interactions used
and also whether the isovector-scalar $\delta $ meson is included or
not. These results imply that it is of great importance to determine
individually each part of the Lorentz covariant nucleon self-energy
decomposition of $E_{\text{sym}}(\rho )$ and $L(\rho )$ from
experiments (e.g., Dirac phenomenology ) or microscopic calculations
based on nucleon-nucleon interactions derived from scattering phase
shifts (e.g., DBHF). On the other hand, the Lorentz covariant
nucleon self-energies in asymmetric nuclear matter can also be
determined from quantum chromodynamics (QCD) by means of QCD
sum-rule techniques~\cite{Coh91,Dru10}. The general expressions of
the Lorentz covariant nucleon self-energy decomposition of
$E_{\text{sym}}(\rho )$ and $L(\rho )$ presented in this work are
thus very useful for determining the symmetry energy from QCD. These
studies are in progress.

\section*{Acknowledgments}

The authors would like to thank R. Chen, B.A. Li, X.H. Li and C. Xu
for useful discussions. This work was supported in part by the NNSF
of China under Grant Nos. 10975097 and 11135011, the Shanghai
Rising-Star Program under Grant No. 11QH1401100, the ``Shu Guang"
project supported by Shanghai Municipal Education Commission and
Shanghai Education Development Foundation, the Program for Professor
of Special Appointment (Eastern Scholar) at Shanghai Institutions of
Higher Learning, the Science and Technology Commission of Shanghai
Municipality (11DZ2260700), and the National Basic Research Program
of China (973 Program) under Contract No. 2007CB815004.

\end{document}